\documentclass[twocolumn,
showpacs,
preprintnumbers,
amsmath,
amssymb,
amsfonts,
final,
a4paper,
aps,
floatfix,
citeautoscript,
footinbib,
prb,
superscriptaddress]{revtex4}
\usepackage{times}
\usepackage{graphicx}
\usepackage[latin1]{inputenc}

\def\tlse{Laboratoire de Physique Th\'{e}orique, CNRS and
  Universit\'{e} de Toulouse, UPS, (IRSAMC), F-31062 Toulouse, France}
\def\lptms{Univ Paris-Sud, Laboratoire de Physique Th\'eorique et Mod\`eles Statistiques, UMR8626, Orsay F-91405, France; CNRS, Orsay, F-91405, France.}

\begin{document}

\author{Javier Almeida}
\affiliation{\tlse}
\author{Guillaume Roux}
\affiliation{\lptms}
\author{Didier Poilblanc}
\affiliation{\tlse}

\date{\today}

\title{Pair Density Waves in coupled doped two-leg Ladders}


\begin{abstract}
  Motivated by Resonant X-ray scattering experiments in cuprate ladder
  materials showing charge order modulation of period $\lambda=3$ and
  5 at specific hole densities, we investigate models involving the
  electronic t-J ladders and bosonic chains coupled via screened
  Coulomb repulsion. Extensive density matrix renormalization group
  calculations applied to the ladders/chains supplemented by a
  self-consistent mean-field treatment of the inter-ladder/chain
  coupling provide quantitative estimates of the charge order for
  $\lambda=3,4$ and 5. As previously proposed, such patterns
  correspond to the emergence of pair density waves which stem from
  the strong electronic correlations. We comment on the existence of a
  $\lambda=4$ modulation not seen so far in experiment.
\end{abstract}

\maketitle

\begin{figure}[t]
\centering
\includegraphics[width=.75\columnwidth,clip]{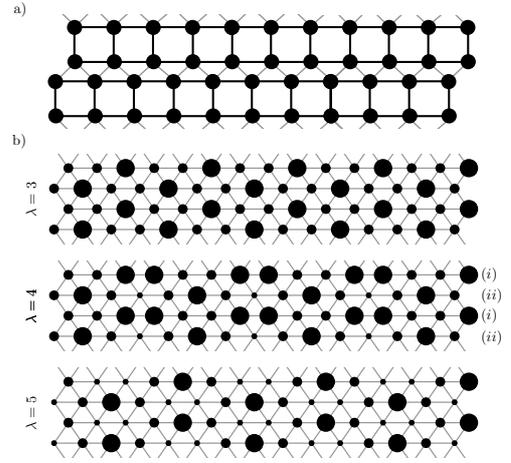}
\caption{(a) Sketch of coupled ladders in a layer of the SCCO
  compound. (b) Proposed CDW modulations for $\lambda=3,4,5$ (as the
  electronic density is identical on both leg, only one is
  represented).}
\label{fig:Sketches}
\end{figure}

It is fascinating that electron correlations alone could lead to
unconventional pairing, eventually leading to superconducting or other
exotic phases such as pair crystals, e.g. in the two-dimensional (2D)
high critical-temperature cuprate superconductors. The quasi
one-dimensional (1D) two-leg ladder material
Sr$_{14-x}$Ca$_{x}$Cu$_{24}$O$_{41}$ (SCCO) is another remarkable
example~\cite{Dagotto1996} of strongly correlated material displaying
both unconventional superconducting and charge-density wave (CDW)
phases. It is a layered compound with intercalated chain and two-leg
ladder subunits. Under doping and high hydrostatic pressure, the
ladder layers exhibit superconductivity~\cite{Uehara1996}. For $x\!=\!0$,
the SCCO is intrinsically hole-doped and showed a charge modulation
within the chain layers~\cite{Fukuda2002}, which later was attributed
to a buckling of the chains~\cite{Smaalen2003}. On the other hand,
another set of experiments revealed evidence of commensurate charge
modulations, with periods $\lambda$ of 3- and 5-lattice spacings
\emph{along the ladders}, which bear their origin in the interactions
between electrons~\cite{Abbamonte2004, Rusydi2006,
  Rusydi2007}. Surprisingly, these experiments did not find
modulations with period $\lambda\!\!=\!\!4$ in contrast to theoretical results
on isolated ladders~\cite{White2002, Roux2007}, hence suggesting that
the frustrated nature of the zig-zag inter-ladder coupling, forming a
trellis lattice (see Fig.~\ref{fig:Sketches}a), has a crucial role.

In this Rapid Communication, we carry out a systematic investigation
of the 2D charge ordering in models of coupled chains of hard-core
bosons (HCB) and electronic ladders. We use a combination of the
density matrix renormalization group~\cite{White1992} (DMRG) technique
to solve the quasi-1D subsystems and a mean-field (MF) treatment of
the screened Coulomb repulsion between them. Even though our main
focus is the physics of doped spin ladders, the simpler hard-core
boson model is also of great interest since (i) it displays similar
density fluctuations and (ii) it is closely related to the ladder
model in the limit of very large exchange rung coupling~\cite{Siller2001}. 
Results from these general models are lastly compared with other possible
interpretations of the experiments.

\emph{Models and their 1D charge fluctuations} -- We start by
recalling the behavior of charge fluctuations in an isolated doped
two-leg ladder that we describe by the isotropic t-J model:
\begin{equation*}
\mathcal{H} = -t \sum_{\langle{i,j}\rangle,\sigma}\mathcal{P} 
[ c^{\dag}_{i,\sigma} c_{j,\sigma} + \text{h.c.}]\mathcal{P} 
+ J \sum_{\langle{i,j}\rangle}[\mathbf{S}_{i}\cdot\mathbf{S}_{j} -
  \frac{1}{4}n_{i}n_{j}]
\end{equation*}
where $\mathcal{P}$ are Gutzwiller projectors, $c_{i,\sigma}$ the
electron annihilation operators with spin $\sigma$ at site $i$, $n_i$
the density operator and $\mathbf{S}_{i}$ the spin operator. The
antiferromagnetic spin exchange $J$ is set to $0.35t$, a typical value
for cuprates. The model has a spin gap that survives to finite doping
$\delta$~\cite{Roux2006}, beyond $\delta=1/3$. Its low-energy physics
is described by a single charge mode $\phi(x)$ governed by a Luttinger
Liquid (LL) effective Hamiltonian~\cite{Giamarchi2004}:
\begin{equation}
\label{eq:Luttinger}
\mathcal{H} = \int \frac{dx}{2\pi} \left[ uK (\pi\Pi(x))^2 + \frac{u}{K}(\nabla\phi(x))^2\right]\;,
\end{equation}
with $\Pi(x)$ the canonically conjugated field. Due to the presence of
a spin gap, the leading charge fluctuations are not the usual $2k_F$
ones but the $4k_F$ ones. In the following, we will denote this
wave-vector $q=2\pi n\equiv 4k_F$, with $n$ the electronic density
($\delta = 1-n = 1/\lambda$). The density fluctuations have a
power-law decay $\langle{n(x)n(0)}\rangle_c \sim \cos(qx) x^{-2K}$
governed by the Luttinger parameter $K$. Depending on $K$, there is a
competition between a $4k_F$-CDW ($K<1/2$) and a $d$-wave
superconducting ($K>1/2$) phase: the phase diagram has been obtained
numerically~\cite{Hayward1995, White2002}. For the commensurate
fillings $n=3/4$ and $n=1/2$, two different kinds of ordered phases
appear~\cite{White2002, Roux2007}. In particular, strong interactions
yield a commensurate CDW for $\lambda=4$ but no signatures of
translational symmetry breaking was observed~\cite{Roux2007} for
$\lambda=3$ or 5.

Furthermore, some insight into the 2D CDW locking will be obtained
using the 1D following HCB t-V model:
\begin{equation*}
\mathcal{H}= -t\sum_{i}[b^\dagger_{i} b_{i+1}+\text{h.c.}] + V\sum_i n_{i}n_{i+1}\;,
\label{Hamiltonian1b}
\end{equation*}
with $b_i$ the hard-core boson annihilation operator and $V$ the
nearest-neighbor repulsion. The choice is motivated by the analogy
between hole pairs and bosons~\cite{Siller2001}. For isotropic ladder,
a more rigorous mapping would also involve additional bosons for
triplets~\cite{Capponi2002}. However, our bosonic model has the same
effective Hamiltonian (\ref{eq:Luttinger}) as the t-J ladder and its
leading density fluctuations are $2k_F$ ones with exponent $2K$. In
terms of the boson density $n=1/\lambda$, we have exactly the same
wave-vector $q = 2\pi n \equiv 2k_F$ as for the ladders. In spite of
this direct analogy, the behavior of $K$ is different in this
model~\cite{Giamarchi2004}.


\emph{Mean-field treatment and CDW patterns} -- We now turn to the
effect of coupling these quasi-1D charge fluctuations via screened
Coulomb repulsion (we neglect other couplings, such as particle or
pair hoppings). According to the lattice structure of SCCO and recent
ab-initio calculations~\cite{Wohlfeld2010}, we consider that the 
main interactions between adjacent ladders are along the grey bonds of
Fig.~\ref{fig:Sketches}a, the magnitude of which is denoted by
$V_{\perp}$. Using a MF approximation, these couplings boil down to a
chemical potential term $-\mu_i n_i$ with
\begin{equation}
\mu_i = -V_{\perp}\sum_{j=\text{neighbor}(i)} \langle{n_j}\rangle\;.
\end{equation}
The local density $\langle{n_j}\rangle$ is then determined
self-consistently. For $\lambda=3$ and $5$, the patterns which
minimize the energy are naturally site-centered CDW shifted by $\pi$
(see Fig.~\ref{fig:Sketches}b) corresponding to
\begin{equation}
\langle{n_j}\rangle = n[1 + \rho_q\cos(q(j-\varphi)) + \rho_{2q}\cos(2qj)]
\label{eq:CDW}
\end{equation}
with $\varphi=0$. Notice that there actually is only one harmonic for
$\lambda=3$ and two for $\lambda=5$ with the particularity that
$2q\equiv q/2$ (hence $2q = 8k_F \equiv 2k_F$ in ladders). In the
limit of small $V_{\perp}$, the MF scheme is actually equivalent to
static RPA as the condition for the order to develop is $1 = z
V_{\perp} \gamma(q) \chi_{1D}(q)$, with $\gamma(q)$ a geometrical
prefactor, $z$ the coordination number and $\chi_{1D}(q)$ the static
charge susceptibility of the isolated quasi-1D system. Classical
configurations with one particle each $\lambda$ site, expected in the
large $V_{\perp}$ limit, are also described by Eq.~(\ref{eq:CDW}) so
that the MF scheme interpolates between the perturbative (RPA) regime
to the non-perturbative (classical) one. In both repulsive models
under study, for which $K<1$, $\chi_{1D}(q)$
diverges~\cite{Giamarchi2004} and the order builds up as soon as
$V_{\perp}$ is branched. Focusing now on the classical limit when
$\lambda=4$, the frustration of the zig-zag couplings makes it
impossible to accommodate two chains with one particle every four
sites, as it happens for the square lattice. Striped configurations
could be realized but we actually found that the most stable pattern
is again two shifted CDW~\footnote{In addition to numerical checks, we
  can argue that in the perturbative regime the $\lambda=4$ staggered
  pattern is stabilized w.r.t. the stripes configuration as the
  interaction energy difference (per site) is $e_{\text{stagg}} -
  e_{\text{int}} \propto -\rho_{q}^2(\sqrt{2}-1)$. In the classical
  limit where there is one particle each 4 site in (i), we see that
  the interaction energy is zero in the staggered pattern while it is
  positive for the stripes.} of the type (\ref{eq:CDW}) (see
Fig.~\ref{fig:Sketches}b) which can yet have different amplitudes,
contrary to the $\lambda=3,5$ cases. The CDW (i) is bond-centered
($\varphi=1/2$) while the other (ii) is site-centered. Here again, the
RPA argument gives that the order develops at any finite
$V_{\perp}$. Numerical calculations are performed taking one finite
quasi-1D system with open boundary conditions and solved by DMRG,
keeping 800 states~\footnote{The density patterns do not require a lot
  of states to converge, particularly when $V_{\perp}\neq 0$. For
  $V_{\perp}=0$, we checked using up to 1600 states.}, embedded
between two infinite ones displaying the pattern (\ref{eq:CDW}). When
$\lambda=4$, the two patterns (i) and (ii) are solved
simultaneously. In the self-consistent procedure, the $\rho_{q}$'s are
sampled in the bulk and the convergence criteria is to have a relative
energy error smaller than $10^{-5}$, for the which the density
patterns are converged.


\begin{figure}[t]
\centering
\includegraphics[width=.85\columnwidth,clip]{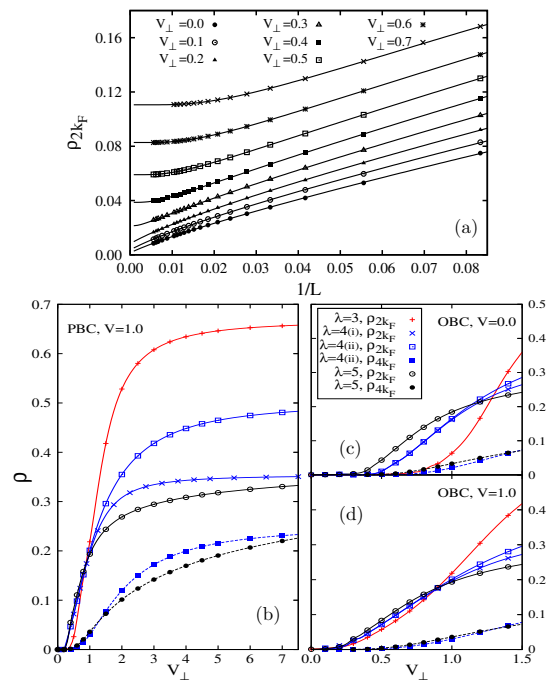}
\caption{(Color online) \emph{HCB model}: (a) Finite size scaling of
  $\rho_{q}$ for $\lambda=3$ and $V=1$. (b) results with PBC showing
  the large $V_{\perp}$ classical limit. (c-d) Extrapolated results
  with OBC showing the opening of the order for $\lambda=3,4,5$ when
  $V=0$ and $V=1$.}
\label{fig:hcb}
\end{figure}

\emph{HCB model} -- We start the discussion with the HCB model. As it
can be solved efficiently, we use both periodic (PBC) and open
boundary conditions (OBC) to check the finite size effects. For PBC, a
finite-size chain does not display an order below a certain value
$V^c_{\perp}(L)$. This can be understood in terms of a correlation
length $\xi$ larger than $L$, making the scaling of the order
parameter with $L$ difficult with PBC. With OBC, there always is some
charge fluctuations at the edges which decay toward the bulk value,
the so-called Friedel oscillations. If the field $\phi$ is not pined,
these oscillations are well fitted by the following ansatz based on LL
theory and symmetries:
\begin{equation}
n(x) = n + A\frac{\cos q(x -(L+1)/2)}{\left[\sin(\pi x/(L+1))\right]^K}
\label{eq:friedel}
\end{equation}
which gives an access to the Luttinger parameter $K$ (the decay
exponent of the $2q$ harmonic is $4K$ so that its contribution is
negligible). When $\phi$ gets pined, the decay of the oscillations are
exponentially suppressed, as $e^{-x/\xi}$. Using OBC, we start the MF
iterations from the $V_{\perp}=0$ density profile that displays
Friedel oscillations and, once convergence is reached, we scale the
bulk value of $\rho_q$ using the following ansatz which interpolates
between the two regimes:
\begin{equation}
\rho_{q}(L)=\rho_{q}^{\infty}+B\;e^{-L/\xi}/L^\alpha \,.
\label{eq:scaling}
\end{equation}
According to the results gathered in Fig.~\ref{fig:hcb}a, such an
ansatz looks reasonable. In Fig.~\ref{fig:hcb}(b-d), one finds the
opening of the order through the behavior of its Fourier components
$\rho_q$ for $\lambda=3,4,5$ both in the non-interacting ($V=0$) and
interacting ($V=1$) regimes. As predicted by RPA, the $q$-component
opens first while the harmonic at $2q$ develops above $V_{\perp}
\gtrsim 0.5$. Remarkably, $\rho_{2q}$ hardly has finite-size
effects. Notice that, for $\lambda=4$, the main Fourier components of
the two inequivalent chains (i) and (ii) have the same magnitude as
one may infer in the perturbative regime of the continuum limit. In
the large $V_{\perp}$ limit, the Fourier components saturate close to
their classical expectations for $\lambda=3$ and 4(i). For
$\lambda=5$, the nearest-neighbor repulsion is not sufficient to pin
the bosons every five sites. We found that such a classical pattern
can be realized by taking into account longer ranged interactions
within the MF approximation (data not shown). In all cases, we
conclude that the patterns of Fig.~\ref{fig:Sketches}b, similar to a
Wigner crystal of bosons, can be realized with a sizable order for
sufficiently large yet \emph{short range} interactions.

\begin{figure}[t]
\centering
\includegraphics[width=.8\columnwidth,clip]{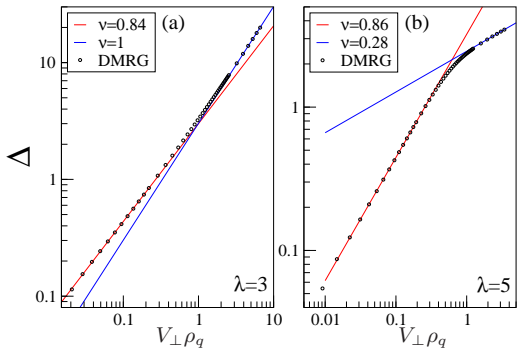}
\caption{(Color online) Scaling of the opening of the gap for
  $\lambda=3$ (a) and $\lambda=5$ (b) in the HCB model with $V=1$
  (extrapolated data).}
\label{fig:scaling}
\end{figure}

Although LL theory cannot give a quantitative prediction for the
magnitude of $\rho_q$, it provides the scaling of the gap associated
with the ordering. The one boson gap is defined as
\begin{equation*}
\Delta(L) = E_0(N+1)+E_0(N-1)-2E_0(N)
\end{equation*}
with $E_0(N)$ the ground-state energy with $N$ bosons. All energies
are computed with the same external $\mu_i$ obtained from the
converged density pattern. In other words, we assume that the
surrounding charge distribution of the neighboring chains is not
perturbed by removing/adding one particle. Gaps are then extrapolated
using $\Delta(L) = \Delta^{\infty} + B e^{-L/\xi}/L$. At the level of
LL theory, the MF coupling yields a sine-Gordon term~\cite{Schulz1983}
which generically reads
\begin{equation}
g \int dx \cos \beta\phi(x)\;.
\label{eq:sinegordon}
\end{equation}
Such a perturbation is relevant when $K<K_c = 8/\beta^2$.  For the
main harmonic $q$, one has $g \propto V_{\perp} \rho_q$ and
$\beta=2$. Thus, we have $K_c=2$ so that when $\rho_q \neq 0$, one is
deep into the massive phase of (\ref{eq:sinegordon}). In this regime,
the gap scales\footnote{In the non-interacting limit where $K=1$, a
  BCS-like opening $\Delta \propto e^{-\text{const.}/V_{\perp}}$ is
  expected at small $V_{\perp}$.} as $\Delta \propto g^{\nu}$ with
$\nu = K_c/2(K_c-K) = 1/(2-K)$. In Fig.~\ref{fig:scaling}a, we report
the scaling of the gap for $V=1$ and $\lambda=3$ (for which there is
no extra harmonic), suggesting an exponent $\nu \simeq 0.84$. A
perturbative estimate of $K$ is $[1 + V/(\pi t \sin(\pi n))]^{-1/2}
\simeq 0.855$ which gives $\nu = 0.87$. A non-perturbative estimate of
$K$ is deduced from fitting the density pattern using
Eq.~(\ref{eq:friedel}): we find $K\simeq 0.78$ and $\nu \simeq
0.82$. The agreement with the sine-Gordon predictions is consequently
very good up to relatively large $V_{\perp}$ (for these particular set
of parameters). Beyond this regime, one recovers $\nu=1$ which is
expected when the $\lambda=3$ pattern saturates: DMRG results nicely
interpolate between the two regimes. In Fig.~\ref{fig:scaling}b,
similar results are found for $\lambda=5$, excepting that the large
$V_{\perp}$ exponent is much smaller ($\nu \simeq 0.28$) because there
is more room in the MF potential for the extra particle. The large
$V_{\perp}$ is thus strongly dependent on the density and the range of
the interactions.

\begin{figure}[t]
\centering
\includegraphics[width=.75\columnwidth,clip]{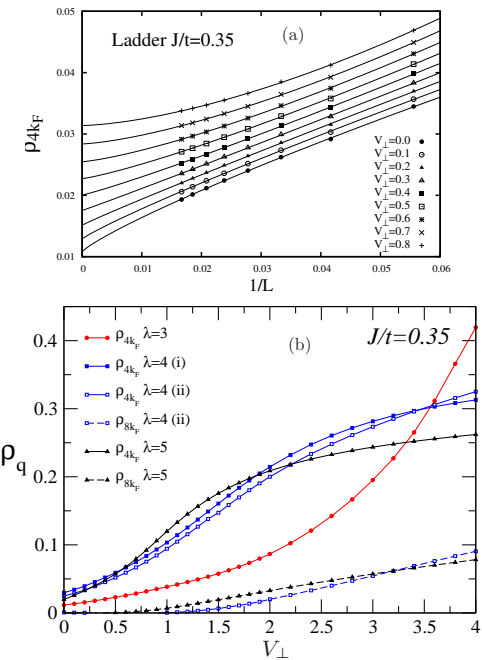}
\caption{(Color online) \emph{tJ model}: (a) finite-size
  extrapolations of $\rho_q$ for $\lambda=3$. (b) Opening of the
  Fourier component with the inter-ladder coupling $V_{\perp}$
  (extrapolated data for $\lambda=3$ and 5). The $\lambda=4$ data are
  shown for $L=60$ and cannot be extrapolated (see text).}
\label{fig:tJ}
\end{figure}

\emph{t-J ladders} -- We now turn to the more realistic case of doped
ladders. First, we recall that, in an isolated ladder, Umklapp
processes can bring a sine-Gordon term (\ref{eq:sinegordon}) with
$\beta=8$ when $\lambda=4$. Even though the corresponding critical
value $K_c = 1/8$ is very small, signatures~\cite{White2002, Roux2007}
of this 1D crystal has been found for small enough $J/t$. From
Eq.~(\ref{eq:friedel}), we get the following values of $K_{\lambda}$
for $J/t=0.35$ on a system with $L=120$: $K_3\simeq 0.55$, $K_4\simeq
0.42$ and $K_5\simeq 0.58$.  We notice a few qualitative differences
with the HCB model. Firstly, $K$ is smaller for $\lambda=4$ suggesting
that this commensurability will lead to the larger order in the
continuum limit. Secondly, the coordination number is now $z=2$
instead of $z=4$ for HCB. This roughly induces a factor two in the
$V_{\perp}$ required to stabilize a sizable order. In
Fig.~\ref{fig:tJ}a, we give the finite size scaling of the Fourier
amplitudes in the $\lambda=3$ case using (\ref{eq:scaling}). As $K\sim
0.5$, finite-size extrapolations are much harder. For $\lambda=3$ and
5, the fits give $\rho_q^{\infty} \simeq 0.01,0.02$ which is within
numerical accuracy. For $\lambda=4$, there is a finite size effect
which makes the extrapolation impossible: for the (ii) chain, which
pattern is ABCBABC\dots, we must take a ladder with one more rung to
preserve the reflection symmetry about the center. Contrary to the HCB
model (for which there is no issue with the finite size scaling), the
minima of the density are gradually shifted by one site between the
edges and the middle. The consequence is that the pattern
(\ref{eq:CDW}) becomes frustrating at the edges, lowering the overall
order on small systems. We observe in Fig.~\ref{fig:tJ}b that the
order increases slower for $\lambda=3$ than for $\lambda=4,5$. This is
qualitatively different from the HCB model with $V=1$. We may
attribute this difference to the effective behavior of pairs of holes
in the ladder. In fact, hole pairs in isotropic doped ladders can be
viewed as a hole pair resonating with a singlet on a plaquette. The
center of mass of the pair can live on a rung or at the center of the
plaquette so that their effective hard-core bosonic model has twice
the number of sites\cite{Siller2001}. We suggest that the slower
opening of the $\lambda=3$ order is related to the extension of the
hole pairs which favors overlapping at large doping. This (relative)
reduction of the density fluctuations with doping even leads to the
absence\cite{Roux2007} of a CDW order when $n=1/2$
($\lambda=2$). Another consideration is that, at the RPA level, the
geometrical factor $\gamma(q)\propto \cos(\pi/\lambda)$ favors large
$\lambda$. For $V_{\perp}>4$, the $\lambda=3$ curve eventually reaches
the classical expectation. In addition to this feature, hole pairs
have effective interactions at longer range\cite{Siller2001} which
show up in the low $K$ that can be achieved. This should favor
crystallization at small doping and we indeed see that the $\lambda=5$
and 4 curves are comparable.

\emph{Conclusion} -- The emergence of charge ordering in a 2D array of
two-leg ladders with a zig-zag coupling mimicking the crystallographic
structure of the ladder planes of SCCO has been studied using a MF
approximation to treat the screened Coulomb interaction. Our numerical
approach provides quantitative predictions for the order parameter of
charge modulations with period $\lambda =3,4,5$. A good agreement with
LL theory is found in the perturbative regime before reaching the
classical limit. The analogy between the results for the two models
suggests that such structures can be interpreted physically as pair
density waves, a localization of Cooper pairs with no superconducting
coherence~\cite{Chen2004}, spatially organized as in a Wigner crystal
of pairs~\cite{Rusydi2006}. We note that only the $\lambda =3$ and 5
charge order have been found experimentally while our minimal model
still supports a sizable order for the $\lambda=4$ modulation.  An
Hartree-Fock approach to a more sophisticated
model~\cite{Wohlfeld2007} suggested a relative reduction of the
$\lambda=4$ order but there is no clear suppression. Another proposal
was to take into account a magnetic ring exchange~\cite{Rusydi2007}
that lowers the $\lambda=4$ charge order. However, according to the
behavior of the pairing energy with doping and ring exchange in the
t-J model~\cite{Roux2005}, the $\lambda=5$ modulation should also be
suppressed. A possible explanation could be that the experimental
set-up was not able to probe the peculiar type of ordering of
Fig.~\ref{fig:Sketches}b which involves two inequivalent
ladders. Lastly, the $\lambda=4$ pattern could be more sensitive to impurities, 
magnetic exchange, single particle hopping or pair tunneling
between ladders leading to a competition with superconducting phases


\begin{thebibliography}{10}

\bibitem{Dagotto1996}
E.~Dagotto and T.~M. Rice, Science {\bf 271}, 618 (1996).

\bibitem{Uehara1996}
M.~Uehara {\em et~al.}, Journal of the Physical Society of Japan
{\bf 65}, 2764 (1996);
H.~E. K.M.~Kojima, N.~Motoyama and S.~Uchida, J. Electron
Spectrosc. Relat. Phenom. {\bf 117-118}, 237 (2001).

\bibitem{Fukuda2002}
T.~Fukuda, J.~Mizuki, and M.~Matsuda, Phys. Rev. B {\bf 66}, 012104
(2002).

\bibitem{Smaalen2003}
S.~van Smaalen, Phys. Rev. B {\bf 67}, 026101 (2003).

\bibitem{Abbamonte2004}
P.~Abbamonte {\em et~al.}, Nature {\bf 431}, 1078 (2004).

\bibitem{Rusydi2006}
A.~Rusydi {\em et~al.}, Phys. Rev. Lett. {\bf 97}, 016403 (2006);
A.~Rusydi {\em et~al.}, Phys. Rev. Lett. {\bf 100}, 036403
(2008).

\bibitem{Rusydi2007}
A.~Rusydi {\em et~al.}, Phys. Rev. B {\bf 75}, 104510 (2007).

\bibitem{White2002}
S.~R. White, I.~Affleck, and D.~J. Scalapino, Phys. Rev. B {\bf 65},
165122 (2002).

\bibitem{Roux2007}
G.~Roux, E.~Orignac, S.~R. White, and D.~Poilblanc, Phys. Rev. B {\bf
  76}, 195105 (2007).

\bibitem{White1992}
S.~R. White, Phys. Rev. Lett. {\bf 69}, 2863 (1992);
U.~Schollw\"ock, Rev. Mod. Phys. {\bf 77}, 259 (2005).

\bibitem{Siller2001}
T.~Siller, M.~Troyer, T.~M. Rice, and S.~R. White, Phys. Rev. B {\bf 63}, 195106 (2001).

\bibitem{Roux2006}
G.~Roux, S.~R. White, S.~Capponi, and D.~Poilblanc,
Phys. Rev. Lett. {\bf 97}, 087207 (2006).

\bibitem{Giamarchi2004}
T.~Giamarchi, {\em Quantum Physics in one Dimension} International
series of monographs on physics Vol. 121 (Oxford University Press,
Oxford, UK, 2004).

\bibitem{Hayward1995}
C.~A. Hayward, D.~Poilblanc, R.~M. Noack, D.~J. Scalapino, and
W.~Hanke, Phys. Rev. Lett. {\bf 75}, 926 (1995).

\bibitem{Capponi2002}
S.~Capponi and D.~Poilblanc, Phys. Rev. B {\bf 66}, 180503(R) (2002).

\bibitem{Wohlfeld2010}
K.~Wohlfeld, A.~M. Ole\'{s}, and G.~A. Sawatzky, Phys. Status Solidi B {\bf 247}, 668 (2010);
\emph{private communication} (2010).

\bibitem{Schulz1983}
H.~J. Schulz and C.~Bourbonnais, Phys. Rev. B {\bf 27}, 5856 (1983);
S.~T. Carr and A.~M. Tsvelik, Phys. Rev. B {\bf 65}, 195121 (2002).

\bibitem{Chen2004}
H.-D. Chen, O.~Vafek, A.~Yazdani, and S.-C. Zhang,
  Phys. Rev. Lett. {\bf 93}, 187002 (2004).

\bibitem{Wohlfeld2007}
K.~Wohlfeld, A.~M. Ole\'{s}, and G.~A. Sawatzky, Phys. Rev. B {\bf
    75}, 180501 (2007).

\bibitem{Roux2005}
G.~Roux, S.~R. White, S.~Capponi, A.~L\"auchli, and D.~Poilblanc,
  Phys. Rev. B {\bf 72}, 014523 (2005).

\end{thebibliography}
\end{document}